\documentclass[prd, 12pt, tightenlines, nofootinbib]{revtex4}

\newif\ifshowtitles
\showtitlestrue

\def\ptitle#1{\ifshowtitles {\it #1}\fi}
\ifshowtitles \def\em{} 
  \fi                   

\catcode`@=11
   \newif\ifRevTexFour \newif\ifclassdef
   \def\ifundefined#1{\expandafter\ifx\csname#1\endcsname\relax}
   \def\rttest{revtex4}
   \ifundefined{class@name} \else \classdeftrue
      \ifx \rttest \class@name \RevTexFourtrue \fi\fi
   \ifRevTexFour
      \usepackage{epsfig}
      \def\bodymatter{\maketitle}
      \def\refcite#1{\cite{#1}}
      \def\eref#1{Eq.~(\ref{#1})}
      \def\fref#1{Fig.~(\ref{#1})}
      
   \fi
\catcode`@=12

\def\figsubcap#1{\par\noindent\centering\footnotesize(#1)}
\def\simgt{\mathrel{\lower2.5pt\vbox{\lineskip=0pt\baselineskip=0pt
           \hbox{$>$}\hbox{$\sim$}}}}
\def\simlt{\mathrel{\lower2.5pt\vbox{\lineskip=0pt\baselineskip=0pt
           \hbox{$<$}\hbox{$\sim$}}}}
\def\simprop{\mathrel{\lower3.0pt\vbox{\lineskip=1.0pt\baselineskip=0pt
             \hbox{$\propto$}\hbox{$\sim$}}}}

\begin{document}

\preprint{MIT-CTP 4489}

\title{QUANTUM FLUCTUATIONS IN COSMOLOGY\\AND HOW THEY LEAD TO A
MULTIVERSE\footnote{Talk given at the 25th Solvay Conference on
Physics, {\it The Theory of the Quantum World,} Brussels, 19--22 October
2011.  Published in the Proceedings, edited by D.~Gross,
M.~Henneaux, and A.~Sevrin (World Scientific, 2013).}}

\author{Alan H. Guth$^{**}$}

\address{Center for Theoretical Physics, Laboratory for Nuclear Science, 
     and Department of Physics,\\
Massachusetts Institute of Technology, Cambridge, MA 02139, USA\\
$^{**}$E-mail: guth@ctp.mit.edu\\
web.mit.edu}

\begin{abstract}
This article discusses density perturbations in inflationary
models, offering a pedagogical description of how these
perturbations are generated by quantum fluctuations in the early
universe.  A key feature of inflation is that that rapid
expansion can stretch microscopic fluctuations to cosmological
proportions.  I discuss also another important conseqence of
quantum fluctuations: the fact that almost all inflationary
models become eternal, so that once inflation starts, it never
stops. 
\end{abstract}


\bodymatter

\section{Introduction}\label{ahg:sec1}

I have been asked to describe quantum fluctuations in cosmology,
which I find a fascinating topic.  It is a dramatic demonstration
that the quantum theory that was developed by studying the
hydrogen atom can be applied on larger and larger scales.  Here
we are applying quantum theory to the universe in its entirety,
at time scales of order $10^{-36}$ second, and it all sounds
incredibly fantastic.  But the shocking thing is that it works,
at least in the sense that it gives answers for important
questions that agree to very good precision with what is actually
measured.  In addition to discussing the density perturbations
that we can detect, however, I want to also discuss another
important aspect of quantum fluctuations:  specifically, quantum
fluctuations in cosmology appear, in almost all our models, to
lead to eternal inflation and an infinite multiverse.  This is a
rather mind-boggling concept, but given our success in
calculating the fluctuations observed in the cosmic microwave
background (CMB), it should make good sense to consider the other
consequences of quantum fluctuations in the early universe. 
Thus, I think it is time to take the multiverse idea seriously,
as a real possibility.  The inhomogeneities that lead to eternal
inflation are nothing more than the long-wavelength tail of the
density perturbations that we see directly in the CMB.

\section{Origin of Density Perturbations During the Inflationary
Era}\label{ahg:sec2}

The idea that quantum fluctuations might be the origin of
structure in the universe goes back at least as far as a 1965
paper by Sakharov~\cite{Sakharov65}. In the context of
inflationary models, the detailed predictions are
model-dependent, but a wide range of simple models give generic
predictions which are in excellent agreement with observations. 
In this section I will give a pedagogical explanation of how
these predictions arise, based on the time-delay formalism that
was used in the paper I wrote with S.-Y.~Pi~\cite{GuthPi82}.
This formalism, which we learned from Stephen Hawking, is the
simplest to understand, and it is completely adequate for the
dominant perturbations in single-field, slow-roll
inflation.\footnote{The original work on density perturbations arising from
scalar-field-driven inflation centered around the Nuffield
Workshop on the Very Early Universe, Cambridge, U.K., June-July
1982.  Four papers came out of that workshop:
Refs.~\refcite{Starobinsky82}, \refcite{GuthPi82},
\refcite{Hawking82}, and \refcite{BST83}.  Ref.~\refcite{BST83}
introduced a formalism significantly more general than the
previous papers.  These papers tracked the perturbations from
their quantum origin through Hubble exit, reheating, and Hubble
reentry.  Earlier Mukhanov and
Chibisov~\cite{Mukhanov-Chibisov81} had revived Sakharov's idea
in a modern context, studying the conformally flat perturbations
generated during the inflationary phase of the Starobinsky
model~\cite{Starobinsky80}.  They developed a method of
quantizing the metric fluctuations, a method more sophisticated
than is needed for the simpler models of
Refs.~\refcite{GuthPi82}--\refcite{BST83}, and gave a formula
(without derivation) for the final spectrum.  For various reasons
the calculations showing how the conformally flat fluctuations
during inflation evolve to the conformally Newtonian fluctuations
after inflation were never published, until the problem was
reconsidered later in Refs.~\refcite{Starobinsky83} and
\refcite{Mukhanov89}.  The precise answer obtained in
Ref.~\refcite{Mukhanov-Chibisov81}, $Q(k) = \sqrt{24 \pi G} M \Bigl(1 + {1
\over 2} \ln (H/k)\Bigr)$, has not (to my knowledge) been
confirmed in any modern paper.  However, the fact that $Q(k)$ is
proportional to $\ln({\rm const}/k)$ has been confirmed, showing
that the 1981 paper by Mukhanov and Chibisov did correctly
calculate what we now call $n_s$ (as was pointed out in
Ref.~\refcite{Hinshaw12}).}
More sophisticated approaches are needed, however, to study
multifield models or models that violate the slow-roll
approximation, or to study extremely subdominant effects in
single-field, slow-roll models.  Even for multifield inflation,
however, some of the simplicity of the time-delay formalism can
be maintained by the use of the so-called $\delta N$
formalism~\cite{SasakiStewart96, BTW06}. There are a number of
reviews~\cite{MalikWands09, BTW06, MFB92} and
textbooks~\cite{LythLiddle09, Weinberg08, Mukhanov05, Dodelson03}
that give a much more thorough discussion of density
perturbations in inflationary models than is appropriate here.

\begin{figure}[t]
\begin{centering}
 \parbox{2.1in}{\epsfig{figure=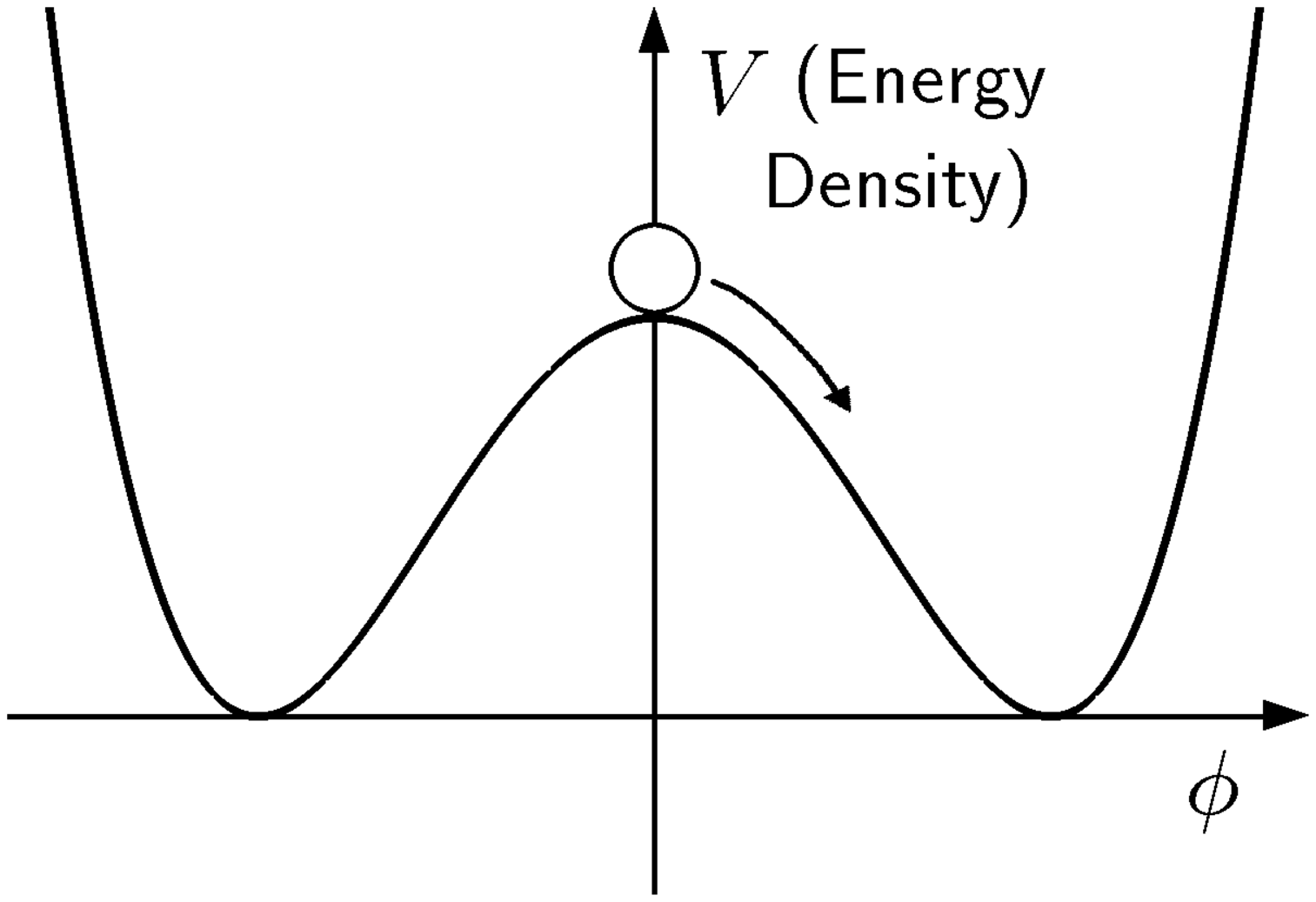,width=2in}
 \figsubcap{a}}
 \hspace*{4pt}
 \parbox{2.1in}{\epsfig{figure=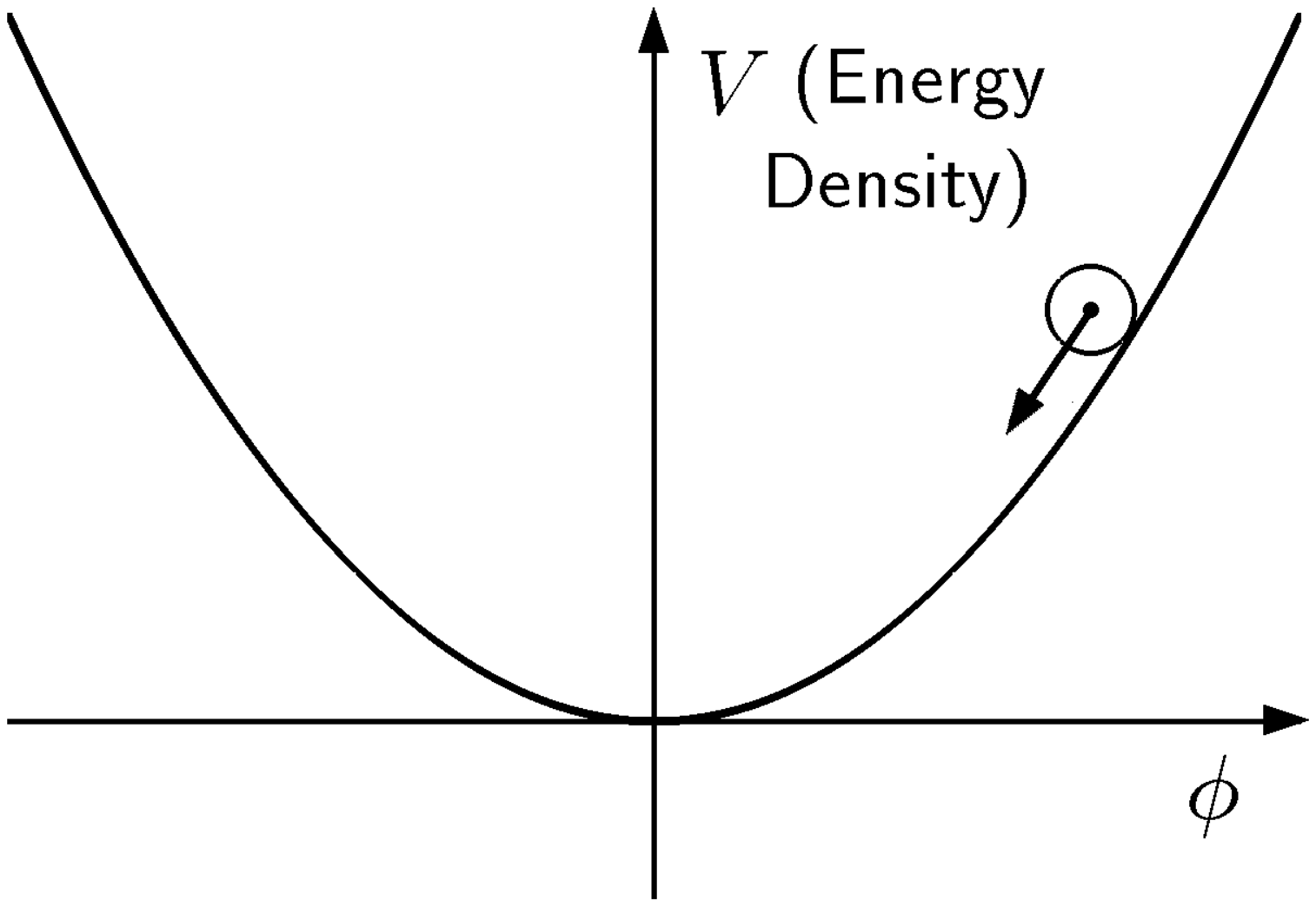,width=2in}
 \figsubcap{b}}
 \caption{(a) Potential energy function for new inflation.  (b)
    Potential energy function for chaotic inflation.}
\label{ahg:fig1}
\end{centering}
\end{figure}

Inflation~\cite{Guth81, Linde82, AlbrechtSteinhardt82} takes place
when a scalar field has a large potential energy density.  A
straightforward application of Noether's theorem~\cite{Noether18}
gives the energy-momentum tensor of a canonically normalized
scalar field as
\begin{equation}
  T^{\mu\nu} = p g^{\mu\nu} + (p + \rho) u^\mu u^\nu \ ,
\label{ahg:eq1}
\end{equation}
where
\begin{eqnarray}
  \rho &=& -{1 \over 2} g^{\mu\nu} \partial_\mu \varphi
     \partial_\nu \varphi + V(\varphi) \ ,\label{ahg:eq2} \\
  p &=& -{1 \over 2} g^{\mu\nu} \partial_\mu \varphi
     \partial_\nu \varphi - V(\varphi) \ ,\label{ahg:eq3} \\
  u^\mu &=& \left( - g^{\lambda\sigma} \partial_\lambda \varphi \,
     \partial_\sigma \varphi \right)^{-1/2} g^{\mu\rho}
     \partial_\rho \varphi \ ,\label{ahg:eq4}
\end{eqnarray}
where $\partial_\mu \equiv \partial/\partial x^\mu$. So, as long
as the energy of the state is dominated by $V(\varphi)$,
\eref{ahg:eq3} guarantees that the pressure is large and
negative.  Einstein's equations imply that negative pressure
creates repulsive gravity, so any state whose energy is dominated
by the potential energy of a scalar field will drive inflation. 
There are two basic scenarios --- one where $\varphi$ starts at
the top of a hill (new inflation~\cite{Linde82,
AlbrechtSteinhardt82}), and one where it starts high on a hill
and rolls down (chaotic inflation~\cite{Linde83}); see
\fref{ahg:fig1}.  Either scenario is successful, and for
the density perturbation calculation we can treat them at the
same time.  We use comoving coordinates, with a background metric
describing a flat Friedmann-Robertson-Walker (FRW) universe:
\begin{equation}
  d s^2 = - d t^2 + a^2(t) \, \delta_{ij} d x^i \, d x^j \ ,
\label{ahg:eq5}
\end{equation}
where $a(t)$ is the scale factor.  Objects moving with the
expansion of the universe are at rest in this coordinate system,
with the expansion described solely by $a(t)$: when $a(t)$
doubles, all the distances in the universe double.  In this
metric the Klein-Gordon equation for a scalar field is given by
\begin{equation}
  \ddot \varphi + 3 H(\varphi,\dot \varphi) \dot \varphi - {1
     \over a^2(t)} \nabla^2 \varphi = - {\partial V(\varphi)
     \over \partial \varphi} \ ,
\label{ahg:eq6}
\end{equation}
where an overdot indicates differentiation with respect to time
$t$, and $\nabla^2$ is the Laplacian, $\sum_i \partial^2 /
\partial (x^i)^2$, with respect to the coordinates $x^i$.  The
equation is identical to the Klein-Gordon equation in Minkowski
space, except that there is a drag term, $3 H \dot \varphi$,
which can be expected, since the energy density must fall if the
universe is expanding.  In addition, each spatial gradient is
modified by $1/a(t)$, which converts the derivative to the
current scale of spatial distance.  The Hubble expansion rate $H
\equiv \dot a / a$ is given by the Friedmann equation for a flat
universe,
\begin{equation}
  H^2 = {8 \pi \over 3} G \left( {1 \over 2} \dot \varphi^2 +
     V(\varphi) \right) \ .
\label{ahg:eq7}
\end{equation}
In this language, the repulsive effect of the negative pressure
that was mentioned above can be seen in the equation for the
acceleration of the expansion,
\begin{equation}
  \ddot a = - {4 \pi \over 3} G (\rho + 3 p) a \ .
\label{ahg:eq8}
\end{equation}
If $p = - \rho$, as one finds when $V(\varphi)$ dominates, this
equation gives $\ddot a = (8 \pi / 3) G \rho a $.

Using an assumption called the slow-roll approximation, which is
valid for a large range of inflationary models, we can ignore the
$\ddot \varphi$ term of \eref{ahg:eq6} and the $\dot \varphi^2$
term in \eref{ahg:eq7}.  In addition, at sufficiently late times
the Laplacian term can be neglected, since it is suppressed by
$1/a^2(t)$.  We are then left with a very simple differential
equation, 
\begin{equation}
  3 H(\varphi) \dot \varphi = - {\partial V \over \partial \varphi} \ ,
\label{ahg:eq9}
\end{equation}
which has a one-parameter class of solutions.  That one parameter
is itself trivial --- it is a time offset.  Given one solution
$\varphi_0(t)$, the general solution can be written as
$\varphi_0(t-\delta t)$, where $\delta t$ is independent of $t$. 
Since the differential equation (\ref{ahg:eq9}) has no spatial
derivatives, $\delta t$ can depend on position, so the most
general solution can be written as
\begin{equation}
  \varphi(\vec x, t) = \varphi_0\bigl(t - \delta t(\vec x)\bigr) \ .
\label{ahg:eq10}
\end{equation}
Since we are interested in developing a first order perturbation
theory, we can expand about $\varphi_0(t)$, 
\begin{equation}
  \varphi(\vec x, t) \equiv \varphi_0(t) + \delta \varphi(\vec x,
     t) = \varphi_0(t) - \dot \varphi_0(t) \, \delta t(\vec x) \ ,
\label{ahg:eq11}
\end{equation}
so 
\begin{equation}
  \delta t(\vec x) = - {\delta \varphi(\vec x,t) \over \dot
     \varphi_0(t)} \ .
\label{ahg:eq12}
\end{equation}
Even though the numerator and denominator of the above expression
both depend on time, the quotient does not.  Thus at late times
(within the inflationary era) --- times late enough for
\eref{ahg:eq9} to be accurate --- the nonuniformities of the
rolling scalar field are completely characterized by a
time-independent time delay.%
\footnote{\label{ahg:footb}%
The description of the perturbations at late times by a
time-independent time delay $\delta t(\vec x)$ is in fact much
more robust than the approximation that $\ddot \varphi$ can be
neglected.  It is a consequence of the Hubble drag term, and will
hold at sufficiently late times in any single-field model for
which the slow-roll approximation is valid for more than a few
$e$-folds.  To see this, consider \eref{ahg:eq6}, with $H$ taken
to be an arbitrary function of $\varphi$ and $\dot \varphi$.  We
will neglect the Laplacian term, since it is suppressed by
$1/a^2(t)$, and we are interested in late times.  Then, for each
value of $\vec x$ there is a two-parameter class of solutions to
this second order ordinary differential equation.  To see the
effect of the damping, suppose that we know the unperturbed
solution, $\varphi_0(t)$, and a nearby solution, $\varphi_0(t) +
\delta \varphi(t)$, where $\delta \varphi(t)$ is to be treated to
first order.  $\delta \varphi$ can depend on $\vec x$, but we
suppress the argument because we consider one value of $\vec x$
at a time.  We then find that $\delta \varphi(t)$ and $\dot
\varphi_0(t)$ obey the same differential equation.  If we
construct the Wronskian $W(t) \equiv \dot \varphi_0 \, \delta
\dot \varphi - \ddot \varphi_0 \, \delta \varphi$, we find that
$$
  \dot W = - 3 \left( H + {\partial H \over \partial \dot
     \varphi} \, \dot \varphi_0 \right) W \ ,
$$
the solution to which is
$$
  W(t) = W_0 \exp \left\{ - 3 \int_{t_0}^t \, d t \left( H +
     {\partial H \over \partial \dot \varphi} \, \dot \varphi_0
     \right) \right\} \ .
$$
Thus $W(t)$ falls off roughly as $e^{- 3 H t}$ or faster ($\dot
\varphi_0 \partial H / \partial \dot \varphi > 0$), and so can be
neglected after just a few $e$-folds of expansion.  Then note
that
$$
   {d \over d t} \left( {\delta \varphi \over \dot \varphi_0}
     \right) = {W(t) \over \dot \varphi_0^2} \ ,
$$
while in the slow-roll regime $\dot \varphi_0^2$ is approximately
constant --- from \eref{ahg:eq9} one can show that the fractional
change in $\dot \varphi_0^2$ during one Hubble time ($H^{-1}$) is
approximately $2(\epsilon-\eta)$, as defined in
Eqs.~(\ref{ahg:eq14}) and (\ref{ahg:eq15}).  Thus the time
derivative of the ratio $\delta \varphi / \dot \varphi_0$ falls
off as $e^{-3 H t}$ or faster, implying that the time delay
rapidly approaches a fixed value.  The time-delay description
remains accurate throughout the reheating process, even though
the slow-roll conditions will generally fail badly at the end of
inflation, when the scalar field starts to oscillate about the
bottom of the potential well.  The time delay is maintained
because $\delta \varphi$ and $\dot \varphi_0$ continue to obey
the same linear differential equation.  Thus if $\delta \dot
\varphi /\delta \varphi = \ddot \varphi_0/\dot \varphi_0$ at the
end of the slow roll period, then $\delta \varphi$ will remain
proportional to $\dot \varphi_0$ for all later times.  This
argument, which generalizes an argument in
Ref.~\refcite{GuthPi82}, is in contradiction with
Ref.~\refcite{Wang97}, where it is argued that the time delay
persists to the end of inflation only under very stringent
assumptions about the potential.  The argument of
Ref.~\refcite{Wang97}, however, is really a discussion of the
validity of \eref{ahg:eq9}, but we have seen that the time delay
is preserved even when \eref{ahg:eq9} fails.} 
It is useful to define a dimensionless measure of the time delay,
\begin{equation}
  \delta N = H \delta t \ ,
\label{ahg:eq13}
\end{equation}
which can be interpreted as the number of $e$-folds of inflation
by which the field is advanced or retarded.

To justify the slow-roll approximation, we must adopt
restrictions on the form of the potential energy function
$V(\varphi)$.  The slow-roll approximation is equivalent to
saying that the field $\varphi$ evolves approximately at the
drag-force limited velocity, where the drag force equals the
applied force, with inertia playing only a negligible role. 
(This would be called the terminal velocity, except that it can
change slowly with time.)  From the first two terms of
\eref{ahg:eq6} one can see that the velocity approaches the
drag-limited value with a time constant of order $H^{-1}$.  Thus,
for the field to evolve at the drag-limited velocity, it is
essential that neither the drag coefficient nor the applied force
changes significantly during a time of order $H^{-1}$.  Thus we
want to insist that $H^{-1} |\dot H| \ll H$, and that $H^{-1} |
(\partial^2 V / \partial \varphi^2) \dot \varphi| \ll |\partial V
/ \partial \varphi| $. Using \eref{ahg:eq9} to approximate $\dot
\varphi$, these two conditions can be expressed in terms of the
two slow-roll parameters~\cite{LythLiddle09}
\begin{eqnarray}
   \epsilon &\equiv& {1 \over 16 \pi G} \left( {V' \over V}
     \right)^2 \approx - {\dot H  \over H^2} \ , \quad 0 <
     \epsilon \ll 1 \ ,
   \label{ahg:eq14} \\
   \eta &\equiv& {1 \over 8 \pi G} {V'' \over V} \approx  - {V'' \dot
     \varphi \over  H V'} \approx \epsilon - {\ddot H \over 2 H
     \dot H}\ , \quad |\eta| \ll 1 \ ,
   \label{ahg:eq15}
\end{eqnarray}
where a prime denotes a derivative with respect to $\varphi$.
Note that these slow-roll conditions do not by themselves
guarantee that $\varphi$ will evolve at drag-limited velocity,
because a large initial velocity will take time before it
approaches the drag-limited value.  But the slow-roll conditions
do guarantee that for times long compared to $H^{-1}$, $\varphi$
will evolve at very nearly the drag-limited velocity.

\begin{figure}[t]
\begin{centering}
\epsfig{file=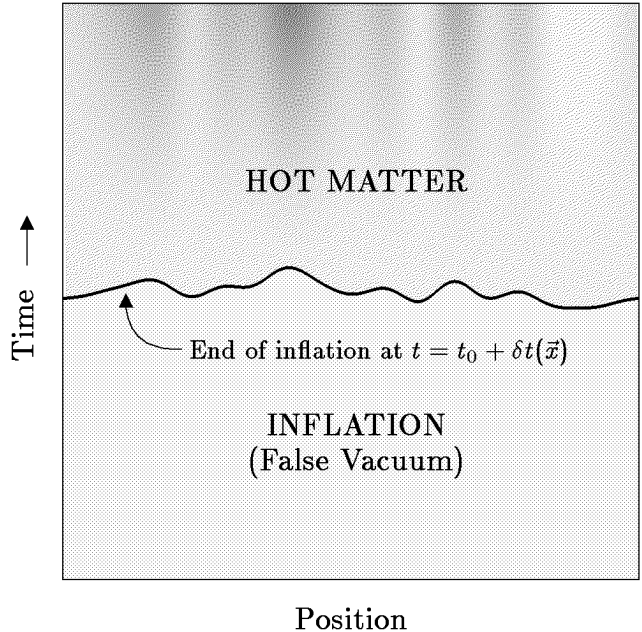}
\caption{Schematic illustration of our approximations: 1)
before the end of inflation, shown as a wiggly line, the
spacetime is the unperturbed, exponentially expanding flat (de
Sitter) space corresponding to the ``false vacuum'' state of the
scalar field; 2) inflation ends on a sharp line, at which the
matter is immediately transformed into thermal radiation.}
\end{centering}
\label{ahg:fig2}
\end{figure}

To proceed, I will make two approximations that will simplify the
problem enormously, but which are nonetheless extremely accurate
for single-field slow-roll inflation.  First, we will neglect all
perturbations of the metric until the time when inflation ends. 
That is, until inflation ends we treat the scalar field as a
quantum field in a fixed de Sitter space background.  Thus, we
will be ignoring the fluctuations in the energy-momentum tensor
of the scalar field, since we are not allowing them to perturb
the metric.  However, we will calculate the fluctuations in the
scalar field itself, as described by the time delay $\delta
t(\vec x)$.  Since the scalar field is driving the inflation, the
time delay $\delta t(\vec x)$ measures the variation in the time
at which inflation ends at different places in space.  The amount
of energy that is released at the end of inflation is much larger
than the energy-momentum tensor fluctuations during inflation, so
the spatial variation of the timing of this energy release
becomes the dominant source of the density perturbations that
persist at later times.  To describe this release of energy, we
make our second approximation.  We will treat the ending of
inflation as instantaneous.  We will assume that the potential
energy of the inflaton field is converted instantaneously into
the thermal radiation of effectively massless particles,
beginning the radiation-dominated era of cosmological history.

I will give heuristic justifications for these approximations,
but I am not aware of a more rigorous justification that can be
explained without developing an understanding of what happens
when these approximations are avoided, and then the problem
requires a much more detailed analysis.  Such analyses have of
course been done and are even described in textbooks.  The answer
that we will obtain agrees with the textbooks~\cite{LythLiddle09,
Weinberg08, Dodelson03}, for the single-field slow-roll case,
down to the last factor of $\sqrt{\pi}$.  While the textbooks
corroborate the answer that we will obtain, the methods are
sufficiently different so that very little light is shed on the
approximations described here.  In a future
publication\cite{Guth12}, I will attempt a more detailed
justification.

For a given theory we can calculate $\varphi_0(t)$ by solving an
ordinary differential equation, so \eref{ahg:eq12} reduces the
problem of calculating $\delta t(\vec x)$ to that of calculating
the fluctuations of the scalar field, $\delta \varphi(\vec x,t)$. 
This is a problem in quantum field theory, albeit quantum field
theory in curved spacetime.  The calculations closely resemble
the familiar quantum field theory calculations for Minkowski
space, but there is one point in the calculation where cosmology
rears its head.  While Minkowski space has a well-defined vacuum
state, which is the starting point for most calculations, it is
less clear what quantum state should be used to describe the
fields evolving in de Sitter space, where the time dependence
prevents the existence of a conserved total energy.  In principle
the quantum state is determined by the initial conditions for the
universe, about which we know very little.  However, while we do
not know the quantum state of the early universe, there is a very
natural choice, corresponding at least locally to the concept of
a vacuum state.  To understand this choice, recall that we are
interested in an exponentially expanding space, the de Sitter
spacetime of inflation, so to a good approximation $a(t) \propto
e^{H t}$, where $H$ is constant.  If we now treat the inflaton
field $\varphi(\vec x, t)$ as a quantum operator, we can as usual
consider its Fourier transform:
\begin{equation}
  \varphi(\vec x,t) = {1 \over (2 \pi)^3} \int \, d^3 k \,
          e^{i \vec k \cdot \vec x} \tilde \varphi(\vec k,t) \, .
\label{ahg:eq16}
\end{equation}
For a free quantum field theory in Minkowski spacetime, $\tilde
\varphi(\vec k,t)$ would be the sum of an annihilation operator
term for particles of momentum $\vec k$ and a creation operator
term for particles of momentum $-\vec k$, each corresponding to a
de Broglie wavelength $\lambda = 2 \pi/|\vec k|$.  For an FRW
spacetime, since the Fourier transform is defined in terms of the
{\it comoving} coordinates $\vec x$, the physical wavelength for
a mode $\vec k$ is not constant, but is given by
\begin{equation}
  \lambda_{\rm phys}(t) = a(t) {2 \pi \over |\vec k|} \, .
\label{ahg:eq17}
\end{equation}
In other words, each mode is stretched as the universe expands. 
Thus, if we follow any mode backwards in time, it will have a
shorter and shorter wavelength and a higher and higher frequency. 
The Hubble expansion rate $H$ is approximately constant during
inflation, so at very early times $H$ is very small compared to
the frequency, and hence is negligible.  Thus, any given mode
behaves at asymptotically early times exactly like a mode in
Minkowski space, so the ``natural'' initial state is to simply
start each mode in its Minkowski vacuum state in the asymptotic
past.  This is called the Bunch--Davies
vacuum~\cite{BunchDavies78}, and it is identical to what is also
called the Gibbons--Hawking vacuum~\cite{GibbonsHawking77}.
Gibbons and Hawking developed their description of the vacuum in
a completely different formalism, based on the symmetries of the
de Sitter spacetime (\eref{ahg:eq5} with $a(t) \propto e^{H t}$),
but the two vacuum states are identical, as one would hope.  The
Bunch--Davies / Gibbons--Hawking vacuum is taken as the starting
point for all standard calculations of density perturbations.

To discuss the spectrum of fluctuations of a spatially varying
quantity such as $\delta \varphi(\vec x,t)$, which is assumed to
be statistically homogeneous and isotropic, cosmologists define
a power spectrum $P_{\,\delta \varphi}(k,t)$ by%
\footnote{Conventions vary, but here we follow the conventions of
Refs.~\refcite{LythLiddle09} and \refcite{Dodelson03}.  The
quantity $\Delta f(\vec k)$ defined in Ref.~\refcite{GuthPi82} is
related by $\Delta f(\vec k)^2 = k^3 P_f(k)/(2 \pi)^3$. 
Another common normalization, called ${\cal P}(k)$ in
Ref.~\refcite{LythLiddle09} and $\Delta^2(k)$ in
Ref.~\refcite{Dodelson03}, is given by ${\cal P}(k) =
\Delta^2(k) = k^3 P(k) / 2 \pi^2$, so $\Delta f(\vec k)^2
= {\cal P}(k)/4 \pi$.  According to Ref.~\refcite{LythLiddle09},
$P(k)$ and ${\cal P}(k)$ are both called {\it the spectrum}.  In
the context of the curvature perturbation ${\cal R}$, to be
defined below, Ref.~\refcite{LiddleLyth00} 
defines yet another normalization that remains in common use,
$\delta_H \equiv {2 \over 5} {\cal P}_{\cal R}$.}
\begin{equation}
  \langle \delta \tilde \varphi (\vec k,t) \, \delta \tilde
     \varphi(\vec k',t) \rangle \equiv (2 \pi)^3 P_{\,\delta
     \varphi}(k,t) \, \delta^{(3)}(\vec k + \vec k') \ ,
\label{ahg:eq18}
\end{equation}
or equivalently
\begin{equation}
  \langle \delta \varphi(\vec x,t) \, \delta \varphi(\vec y,t)
     \rangle = {1 \over (2 \pi)^3} \int \, d^3 k e^{i \vec k
     \cdot (\vec x - \vec y)} P_{\,\delta \varphi} (k,t) \ .
\label{ahg:eq19}
\end{equation}
When $\delta \varphi(\vec x,t)$ is a quantum field operator, the
power spectrum is nothing more than the equal-time propagator,
which can be calculated straightforwardly once the vacuum is
specified, as described in the previous paragraph.  From
\eref{ahg:eq6}, $\delta \tilde \varphi (\vec k, t)$ can be seen
to obey the equation
\begin{equation}
  \delta \ddot {\tilde \varphi} + 3 H \delta \dot {\tilde
     \varphi} + {k^2 \over a^2} \delta \tilde \varphi = -
     {\partial^2 V \over \partial \varphi^2} \delta \tilde
     \varphi \ .
\label{ahg:eq20}
\end{equation}
To make use of this equation, we consider its behavior around the
time of Hubble exit, $t_{\rm ex}(k)$, when the wavelength is
approximately equal to the Hubble length, defined more precisely
by
\begin{equation}
   {k^2 \over a^2(t_{\rm ex})} = H^2 \ .
\label{ahg:eq21}
\end{equation}
We assume that the slow-roll conditions of Eqs.~(\ref{ahg:eq14})
and (\ref{ahg:eq15}) are valid within several Hubble times
($H^{-1}$) of $t_{\rm ex}$ (but it is okay if they are violated
later during the period of inflation, as described in
footnote~\ref{ahg:footb}). For $t \simgt t_{\rm ex}$, the
$k^2/a^2$ term of \eref{ahg:eq20} becomes insignificant.  From
\eref{ahg:eq15} we see that the right-hand-side of
\eref{ahg:eq20} has magnitude $3 \eta H^2 \, \delta \tilde
\varphi$, where $\eta \ll 1$.  Thus for times up to and including
$t_{\rm ex}(k)$ and a little beyond, we can neglect the
right-hand-side and treat $\delta \varphi(\vec x, t)$ as a free,
massless, minimally coupled field in de Sitter space.  It is then
straightforward to show that%
\footnote{One source where from which this equation can be
deduced is Ref.~\refcite{BunchDavies78}, but note that Eq.~(3.6)
is misprinted, and should read $\psi_k(\eta) = \alpha^{-1}
(\pi/4)^{1/2} \eta^{3/2} H_\nu^{(2)}(k \eta)$.}
\begin{equation}
  P_{\, \delta \varphi}(k,t) = {H^2 \over 2 k^3} \left[
     1 + \left( {k \over a(t) H} \right)^2  \right] \ .
\label{ahg:eq22}
\end{equation}
The time delay should be calculated at a time slightly beyond
$t_{\rm ex}$, say by a few Hubble times, when $k^2/a^2
\ll H^2$, when we can neglect the $k^2/a^2$ terms in both
Eqs.~(\ref{ahg:eq20}) and (\ref{ahg:eq22}).  Using
Eqs.~(\ref{ahg:eq9}), (\ref{ahg:eq12}), and (\ref{ahg:eq14}), one
finds several useful expressions for $P_{\delta N} (k)$:
\begin{equation}
  P_{\delta N} (k) = {H^2 \over \dot \varphi_0^2} P_{\, \delta
     \varphi}(k) = {H^4 \over 2 k^3 \dot \varphi_0^2} = {2 \pi G
     H^2 \over k^3 \epsilon} = {9 \over 2} \left( {8 \pi G \over
     3} \right)^3 {V^3 \over k^3 V'^2} \  . 
\label{ahg:eq23}
\end{equation}
Since the time delay is evaluated a few Hubble times beyond
$t_{\rm ex}(k)$, the quantities $V$, $V'$, and $H$ appearing in
the above expressions should all be evaluated at $\varphi_0(t)$
for $t \approx t_{\rm ex}(k)$.  The distinction between $t_{\rm
ex}(k)$ and a few Hubble times later is important only at higher
order in the slow-roll approximation, since the change in
$\varphi_0$ over a Hubble time is of order $\epsilon$.

Note that the fluctuations in $\delta t$ are inversely
proportional to $\epsilon$, which is a consequence of the
presence of $\dot \varphi_0(t)$ in the denominator of
\eref{ahg:eq12}.  We are now in a position to test the
consistency of the first of our approximations (see
\fref{ahg:fig2}), the approximation of neglecting the
fluctuations in the scalar field energy-momentum tensor.  The
major source of these fluctuations is the fluctuation in
potential energy caused by the fluctuations in $\varphi$.  Thus
$\delta \rho / \rho \approx V' \delta
\varphi/V$, so
\begin{equation}
  P_{\delta \rho/\rho}(k) \approx \left( {V' \over V} \right)^2
     P_{\,\delta \varphi}(k) = {8 \pi G H^2 \epsilon \over k^3}
     \left[ 1 + \left( {k \over a(t) H} \right)^2 \right] \ .
\label{ahg:eq24}
\end{equation}
Thus, the fractional fluctuations in the mass density are of
order $\epsilon^2$ times the dimensionless fluctuations
$P_{\delta N} (k)$ of the time delay, so we expect that we can
neglect the mass density fluctuations if we are interested in
calculating the dominant term in the small $\epsilon$
limit.\footnotemark \ \footnotemark

\footnotetext[5]{One might worry that the second term in square brackets
in \eref{ahg:eq24} becomes large at early times, since $a(t)
\propto e^{H t}$.  But note that the factors of $H$ cancel, so
the term is independent of $H$, and that $k/a(t) = k_{\rm phys}$;
this term is just the short distance divergence that would also
be present in a Minkowski space background.  It leads to
divergences, but those divergences must be canceled by the
prescription for regularizing $T_{\mu\nu}$.  The density
fluctuations of cosmological relevance, which are finite and can
be treated classically at late times, arise from the first term
in square brackets.}
\footnotetext[6]{\label{ahg:footf} The fact that $P_{\delta
\rho/\rho}(k) \propto \epsilon$ is a strong argument to justify
the neglect of metric fluctuations, but some experts may not be
convinced without seeing a more complete formulation in which
metric fluctuations can actually be calculated.  In
Ref.~\refcite{Guth12} I will show how the time-delay formalism
can be embedded in a complete first-order calculation in
synchronous gauge.  (Synchronous gauge is most useful here, since
the time delay is time-independent when measured in proper time. 
Other time coordinates will obscure the underlying simplicity.)
Following the notation of Ref.~\refcite{Weinberg08}, the metric
is written as
\[
   g_{00} = -1, \quad g_{0i} = 0, \quad g_{ij} = a^2(t) \left[
     (1+A) \delta_{ij} + {\partial^2 B \over \partial x^i
     \partial x^j} \right] \, .
\]
Defining a new auxiliary field $\chi$ (not used in
Ref.~\refcite{Weinberg08}) by
\[
   \chi \equiv {1 \over 2} \left( 3 \dot A + \nabla^2 \dot B
     \right) - 3 \left[ {\partial H \over \partial
     \varphi_0} \delta \varphi + {\partial H \over \partial \dot
     \varphi_0} \delta \dot \varphi\right] \, ,
\]
the scalar field equation of motion in this gauge can be written as
\[
   \ddot \varphi + 3 H(\varphi_0,\dot \varphi_0) \delta \dot
     \varphi + 3 \dot \varphi_0 \left[ {\partial H \over \partial
     \varphi_0} \delta \varphi + {\partial H \over \partial \dot
     \varphi_0} \delta \dot \varphi\right] - {1 \over a^2}
     \nabla^2 \delta \varphi + V''(\varphi_0) \delta \varphi +
     \dot \varphi_0 \chi = 0 \, .
\]
\pagebreak
$\chi$ can be found from the source equation
\[
   {\partial \over \partial t} (a^2 H \chi) = \dot H \nabla^2
\left( { \delta \varphi \over \dot \varphi_0} \right) \, .
\]
The full metric can be recovered by using
\[
   \nabla^2 A = 2 a^2 H \chi \, ,
\]
and then using the definition of $\chi$ to find $\dot B$.  Note
that the terms involving partial derivatives of $H$
reproduce to first order the dependence of $H$ on $\varphi$ and
$\dot \varphi$ in \eref{ahg:eq6}, so these ``metric
perturbations'' are taken into account by the time-delay
calculation.  The metric perturbations that are ignored are those
proportional to $\chi$, and they can be seen to be small in
slow-roll inflation.  The source term on the right is
proportional to $\dot H$, which is of order $\epsilon$. The term
enters the scalar field equation of motion with a prefactor of
$\dot \varphi_0$, which contributes another factor of
$\sqrt{\epsilon}$ to the suppression.  At later times near the
end of inflation, when the slow-roll condition might fail badly,
$\chi$ is strongly suppressed by the factor $1/a^2$.  For the
slow-roll solution with $\delta \varphi / \dot \varphi_0 \approx
- \delta t (\vec x)$, the source equation can be solved to give
\[
   \chi(t) \approx {H(t_{\rm ex}) - H(t) \over a^2(t) H(t) }
     \nabla^2 \delta t(\vec x) \, ,
\]
where a constant of integration was chosen so that $\chi(t)
\approx 0$ at Hubble exit.  Thus, this formulation gives a
solid underpinning to the intuitive idea that, at late times,
each region can be treated as an independent Robertson-Walker
universe.  Each independent universe follows essentially the same
history, differing from the other universes by only a time offset.}

\section{Evolution of the Density Perturbations Through the End
of Inflation}\label{ahg:sec3}

Thus, we have reduced the calculation of the density fluctuations
to the schematic description of \fref{ahg:fig2}, with the power
spectrum of the time delay given by \eref{ahg:eq23}.  The role of
quantum theory in this calculation is finished --- it determined
the power spectrum of the time delay.  The rest of the
calculation is general relativity and astrophysics.  Here I will
continue the derivation through the end of inflation, which
carries it far enough to compare with the literature and to at
least qualitatively understand the observational consequences.

To understand the implications of $\delta t(\vec x)$ in the
cosmic evolution as described by \fref{ahg:fig2}, it is
convenient to switch to a new coordinate system in which the end
of inflation happens at $t_0$ everywhere.  Just define $\vec x' =
\vec x$ and $t' = t- \delta t(\vec x')$.  Clearly $d t = d t' +
\partial_i \delta t \, d x^{\prime i}$, so the transformation of
the metric of \eref{ahg:eq5} becomes $d s^2 = g'_{\mu\nu} d
x^{\prime \mu} d x^{\prime \nu}$, where
\begin{equation}
  g'_{00} = -1 \ , \quad g'_{0i} = g'_{i0} = - \partial_i \delta
     t \ , \quad g'_{ij} = a^2\bigl(t' + \delta t(\vec x')\bigr)
     \delta_{ij} \ ,
\label{ahg:eq25}
\end{equation}
with an inverse metric, to first order in $\delta t$, given by
\begin{equation}
  g^{\prime 00} = -1 \ , \quad g^{\prime 0i} = g^{\prime i0} = -
     {1 \over a^2} \partial_i \delta t \ , \quad g^{\prime ij} =
     {1 \over a^2 \bigl( t' + \delta t(\vec x') \bigr)}
     \delta_{ij} \ .
\label{ahg:eq26}
\end{equation}

In the primed coordinates the phase transition happens sharply at
$t' = t_0$, with a sudden change from $p=-\rho \equiv - \rho_{\rm
inf}$ in the inflationary phase to $p_{\rm rad}={1 \over 3}
\rho_{\rm rad}$ in the radiation phase.  During the inflationary
phase the energy-momentum tensor is simply $T_\mu{}^\nu = -
\rho_{\rm inf} \delta_\mu^\nu$, while after the transition it is
given by \eref{ahg:eq1}.  The energy-momentum tensor must be
covariantly conserved, which means that
\begin{equation}
  D_\nu T_\mu{}^\nu{} = \partial_\nu T_\mu{}^\nu + \Gamma^\nu_{\nu
     \lambda} T_\mu{}^\lambda - \Gamma^\lambda_{\nu \mu}
     T_\lambda{}^\nu = 0 \ ,
\label{ahg:eq27}
\end{equation}
where $D_\nu$ denotes a covariant derivative.  The affine
connection coefficients $\Gamma$ will not contain any
$\delta$-functions, so $\partial_\nu T_\mu{}^\nu$ cannot contain
any $\delta$-functions either; thus $T_\mu{}^0$ must be
continuous at $t' = t_0$. This implies that
\begin{eqnarray}
  T_0{}^0{}_{\rm rad} &=& p_{\rm rad} + u^0{}_{\rm rad} \, u_{0,{\rm
     rad}} \, (\rho_{\rm rad} + p_{\rm rad} ) = T_0{}^0{}_{\rm
     inf} = - \rho_{\rm inf} \ ,
     \label{ahg:eq28} \\
  T_i{}^0{}_{\rm rad} &=& (\rho_{\rm rad} + p_{\rm rad}) \, u_{i,{\rm
     rad}} \, u^0{}_{\rm rad} = T_i{}^0{}_{\rm inf} = 0 \ .
     \label{ahg:eq29}
\end{eqnarray}
The second of these equations can only be satisfied if $u_{i,{\rm
rad}} = 0$, because $u^0{}_{\rm rad}$ cannot vanish, as $u^\mu$
is timelike, and $(\rho_{\rm rad} + p_{\rm rad}) = {4 \over 3}
\rho_{\rm rad}$ cannot vanish without violating the first
equation.  Thus, the radiation fluid is necessarily at rest in
the frame of reference in which the phase transition occurs
simultaneously.  Requiring $u^2 = -1$, one finds that
\begin{equation}
  u_{0, {\rm rad}} = -1 \ , \quad u_{i, {\rm rad}} = 0 \ ; \quad
     u^0{}_{\rm rad} = 1 \ , \quad u^i{}_{\rm rad} = {1 \over
     a^2} \partial_i \delta t \ .
\label{ahg:eq30}
\end{equation}
Given the above equation, \eref{ahg:eq28} leads immediately to
$\rho_{\rm rad} = \rho_{\rm inf}$; the energy density is
conserved across the transition.  Since the Einstein equations
are partial differential equations that are second order in time
derivatives, the metric and its first time derivative will be
continuous across $t' = t_0$.  So we have now found all the
information needed to give a well-defined Cauchy problem for the
evolution of the model universe, starting at the beginning of the
radiation-dominated era.  At this point the perturbations of
interest have wavelengths vastly larger than the Hubble length,
but during the subsequent evolution the Hubble length will grow
faster than the perturbation wavelength, so later the
perturbations will come back inside the Hubble length.  The
description of the perturbations through the time of Hubble
reentry was given in
Refs.~\refcite{GuthPi82}--\refcite{BST83}, and in many later
sources, but for present purposes we will stop here.

At this point we can discuss the validity of the second of our
key approximations, the approximation of an instantaneous phase
transition.  The actual transition, during which the scalar field
rolls down the hill in the potential energy diagram and then
oscillates about the minimum and reheats, very likely takes many
Hubble times to complete.  However, we need to keep in mind that
the modes of interest exited the Hubble horizon some 50 or 60
Hubble times before the end of inflation, which means that at the
end of inflation their physical wavelength is of order $e^{50}$
to $e^{60}$, or $10^{21}$ to $10^{26}$, times the Hubble length.
Thus, even if the phase transition takes $10^{10}$ Hubble times,
during this time light would be able to travel less than
$10^{-10}$ wavelengths.  Thus the phase transition is effectively
instantaneous, on the time scale that is relevant for influencing
a wave with the wavelengths under consideration.  (Of course the
reheat energy density that we calculated, $\rho_{\rm rad} =
\rho_{\rm inf}$, was an artificiality of the instantaneous
approximation.  But the calculation can easily be adjusted to
account for a lower reheat energy density, which can be found by
doing a more accurate, homogeneous calculation.  At the end of
inflation we would still obtain, in the primed coordinate system,
a radiation fluid that is at rest, with a uniform energy
density.)

Note that the method used here depended crucially on the
assumption that all parts of the universe would undergo the same
sequence of events, so that the only difference from one place to
another is an overall time offset.  If there were more than one
field for which the quantum fluctuations were relevant, then this
would not be true, since a fluctuation of one field relative to
the other could not be described as an overall time offset. 
Thus, multifield inflation requires a more sophisticated
formalism.

\section{Simplifying the Description}\label{ahg:sec4}

The description we have given so far, specifying the metric and
the matter content, is sufficient to calculate the rest of the
history, but it is rather complicated.  Furthermore, it is
equivalent to many other complicated descriptions, related by
coordinate transformations.  It is therefore very useful to find
a coordinate-invariant way of quantifying the density
perturbations.  One convenient approach is motivated by
considering the Friedmann equation for a universe with spatial
curvature, a universe which might be closed, open, or in the
borderline case, flat:
\begin{equation}
  H^2 = {8 \pi \over 3} G \rho - {k \over a^2} \ .
\label{ahg:eq31}
\end{equation}
Here $k$ is a constant, where positive values describe a closed
universe, and negative values describe an open one.  We are
interested in describing a perturbation of a homogeneous
background universe that is flat, $k=0$.  Thus $\rho(\vec x,t)$
will on average equal $\rho_0(t)$, the value for the background
universe, but it will fluctuate about this average.  $H$ is
normally thought of as part of the global description of the
universe, but it has a locally defined analog given by
\begin{equation}
  H_{\rm loc} \equiv {1 \over 3} D_\mu u^\mu \ ,
\label{ahg:eq32}
\end{equation}
where $D_\mu$ is the covariant derivative and $u^\mu$ is the
fluid velocity.  (Here we will deal only with a radiation fluid,
but in a multicomponent fluid $u^\mu$ can be defined in terms of
the total energy-momentum tensor, as described on p.~225 of
Ref.~\refcite{Weinberg08}.)  We can then define
\begin{equation}
  K(\vec x, t) \equiv a^2(t) \left[ {8 \pi \over 3} G \rho(\vec
     x, t) - H^2_{\rm loc} (\vec x, t) \right] \ .
\label{ahg:eq33}
\end{equation}
This quantity has a property called gauge invariance, which means
that its value for any coordinate point $(\vec x,t)$ is not
changed, to first order in the size of the perturbations, by any
coordinate transformation that is itself of order of the size of
the perturbations.  In this case, the gauge invariance follows
from the fact that, apart from the factor $a^2(t)$ which is
irrelevant for this issue, $K(\vec x, t)$ a quantity that is
coordinate-invariant, and which vanishes for the background
universe.  (Note that coordinate invariance by itself is not
enough; if the quantity varied with time in the background
solution, then its value at $(\vec x,t)$ would change if $t$ were
redefined by a small amount.)  $K$ also has the convenient
property that it remains constant as long as spatial derivatives
can be neglected, because it is exactly conserved for the case of
a homogeneous universe.

We can calculate $K(\vec x, t)$ just after the end of inflation,
using the primed coordinate system but dropping the primes. 
We have $H_{\rm loc} = {1 \over 3} D_\mu u^\mu = {1 \over 3}
g^{-1/2} \partial_\mu ( g^{1/2} u^\mu)$, where $g \equiv - \det
(g_{\mu\nu}) = a^6\bigl(t + \delta t(\vec x)\bigr)$ and $u^\mu$
is given by \eref{ahg:eq30}.  This gives $H_{\rm loc} = H_{\rm
inf} + \nabla^2 \delta t / 3 a^2$, which gives
\begin{equation}
  K = - { 2 \over 3} H_{\rm inf} \nabla^2 \delta t  \, ,
\label{ahg:eq34}
\end{equation}
where $H_{\rm inf}$ is the Hubble expansion rate during
inflation, which we have treated as a constant.%
\footnote{At this point one might worry that the approximation of
instantaneous reheating might be crucial to the answer we
obtained.  If we had used a more realistic picture of slow reheating
which leads to a lower reheat energy density, we might expect
that our method would give $K = - {2 \over 3} H_{\rm reheat}
\nabla^2 \delta t$, which could be much smaller.  Although it is
not obvious, however, the reheat energy density does not affect
$K$, which is conserved for wavelengths large compared to the
Hubble length.  Thus, the value we obtained here could have been
calculated before reheating began, and is equal to the value that
holds long after reheating, whether reheating is fast or slow. 
To understand the evolution of $K$ when $H$ changes, however,
requires a more detailed calculation.  Because of the factor of
$a^2(t)$ in \eref{ahg:eq33}, the value of $K$ is in fact
sensitive to quantities that are suppressed by factors of
$1/a^2$.  To see the conservation of $K$ for long wavelengths,
one needs to include the contribution of the auxiliary field
$\chi$ defined in footnote~\ref{ahg:footf}.  This issue will be
discussed in more detail in Ref.~\refcite{Guth12}.}
As in \eref{ahg:eq6}, $\nabla^2$ denotes the Laplacian operator with
respect to the coordinates $x^i$.  Then, given the power spectrum
of \eref{ahg:eq23} for $\delta N = H
\delta t$, we find a power spectrum for $K$ given by
\begin{equation}
  P_K(k) = {4 k^4 \over 9} P_{\delta N}(k) = {2 k H^4 \over 9
     \dot \varphi_0^2 } = {8 \pi G H^2 k \over 9 \epsilon} = 2
     \left( {8 \pi G \over 3} \right)^3 {V^3 k \over V'^2} \ .
\label{ahg:eq35}
\end{equation}
In the literature $K$ is seldom used, but instead it is much more
common to use a variable called the curvature perturbation ${\cal
R}$, for which the usual definition is somewhat complicated.%
\footnote{See, for example, p.~246 of Ref.~\refcite{Weinberg08}.} 
However, it is shown in Ref.~\refcite{LiddleLyth00} that
\begin{equation}
  K = - {2 \over  3} \nabla^2 {\cal R} \ ,
\label{ahg:eq36}
\end{equation}
so
\begin{equation}
  P_{\cal R} (k) = {9 \over 4 k^4} P_K(k) = P_{\delta N} (k) \ ,
\label{ahg:eq37}
\end{equation}
where $P_{\delta N} (k)$ is given by \eref{ahg:eq23}.  This
answer agrees precisely with the answers obtained in
Refs.~\refcite{LythLiddle09}, \refcite{Weinberg08},
\refcite{Dodelson03}, and \refcite{GuthPi82}.%
\footnote{To compare with Ref.~\refcite{Weinberg08}, note that
$(2 \pi)^3  |{\cal R}^0_q|^2$ in this reference corresponds to
$P_{\cal R} (q)$, and is given on pp. 482 and 491.  To compare
with Ref.~\refcite{LythLiddle09}, note that ${\cal P}_\zeta(k)$,
given on p.~406, corresponds to $k^3 P_{\cal R} (k) / (2 \pi^2)$,
as described on p.~89, and that $M_{\rm Pl}$ is the reduced
Planck mass, $1/\sqrt{8 \pi G}$.  To compare with
Ref.~\refcite{Dodelson03}, use $P_\zeta(k) = P_{\cal R}(k)$,
where $P_\zeta(k)$ is given on p.~170.  As explained in
Ref.~\refcite{Weinberg08}, $\zeta$ and ${\cal R}$ refer to
slightly different quantities, but they agree for wavelengths
long compared to the Hubble length.  To compare with
Ref.~\refcite{GuthPi82}, note that $S=K/(a^2H^2)$, and that the
equation for $S(t'=0)$ describes the conditions just after the
end of inflation, with the scale factor $R(t)=e^{\chi t}$, $\chi
= H_{\rm inf}$.  The quantum fluctuations are quantified in this
paper by $\delta t = \delta \varphi/\dot \varphi_0$, with $\Delta
\varphi^2 = k^3 P_{\, \delta \varphi}(k)/(2 \pi)^3 = H^2/(16
\pi^3)$, in agreement with \eref{ahg:eq22} of the current paper.}

The WMAP seven-year paper~\cite{Komatsu10} quotes ${\cal P}_{\cal
R} (k_0) = (2.43 \pm 0.09) \times 10^{-9}$, where $k_0 = 0.002\
\hbox{Mpc}^{-1}$ and ${\cal P}_{\cal R} (k) = k^3 P_{\cal R}(k) /
(2 \pi^2)$.  (The WMAP papers use $\Delta_{\cal R}^2(k)$ for
${\cal P}_{\cal R} (k)$.)  If these fluctuations come from
single-field slow-roll inflation, we can conclude from
Eqs.~(\ref{ahg:eq37}) and (\ref{ahg:eq23}) that at the time of
Hubble exit,
\begin{equation}
  {V^{3/2} \over M_{\rm Pl}^3 \, V'} = 5.36 \times 10^{-4} \ ,
\label{ahg:eq38}
\end{equation}
where $M_{\rm Pl} = 1/\sqrt{8 \pi G} \approx 2.44 \times 10^{18}$
GeV is the reduced Planck mass.

\section{Deducing the Consequences, Comparing with
Observation}\label{ahg:sec5}

From Eqs.~(\ref{ahg:eq37}) and (\ref{ahg:eq23}) we can also
deduce how the intensity of the fluctuations varies with $k$, a
relation which is parameterized by the scalar spectral index $n_s(k)$,
defined by
\begin{equation}
  n_s-1 = {d \ln {\cal P}_{\cal R} (k) \over d \ln k} \ .
\label{ahg:eq39}
\end{equation}
To evaluate this expression for slow-roll inflation, we use the
last expression in \eref{ahg:eq23} for $P_{\cal R}(k)$; 
we recall that the expression is to be evaluated at $t_{\rm
ex}$ (or some fixed number of Hubble times later), and find that
\eref{ahg:eq21} leads to $d \ln k / d t_{\rm ex} = H$.  Then using
\eref{ahg:eq9} to write $\dot \varphi_0 = - V'/(3 H)$, these
equations can be combined to give~\cite{LiddleLyth92}
\begin{equation}
  n_s - 1 = - {V' \over 8 \pi G V} {d \ln (V^3/V'^2) \over d
     \varphi} = - 6 \epsilon + 2 \eta  \approx 4 {\dot H \over
     H^2} - {\ddot H \over H \dot H} \ .
\label{ahg:eq40}
\end{equation}
The WMAP seven-year paper~\cite{Komatsu10} quotes $n_s - 1 = -0.032
\pm 0.012$.  This result suggests that the slow roll parameters
are indeed quite small, and furthermore they have very plausible
values.  The time of Hubble exit is typically of order 60 Hubble
times before the end of inflation, depending mainly on the reheat
temperature, which means that the natural time scale of variation
is of order $60 H^{-1}$.  If each time derivative in the
right-hand expressions of Eqs.~(\ref{ahg:eq14}) and
(\ref{ahg:eq15}) is replaced by a factor of $H/60$, one sees that
the slow roll parameters are plausibly of order 1/60.

The case $n_s = 1$ is called scale-invariant, because it means
that ${\cal P}_{\cal R}(k)$ is independent of $k$; that is, each
mode has the same strength, at the time of Hubble exit, as any
other mode.  Since ${\cal P}_{\cal R}(k)$ is constant while the
wavelength is long compared to the Hubble length, all modes also
have the same strength at the time of Hubble reentry. 
Single-field inflation produces density fluctuations that are
approximately scale-invariant, because all the modes that are
visible today passed through Hubble exit during a small interval 
of time during inflation, so the conditions under which they were
generated were very similar.

In addition to the nearly scale-invariant spectrum that we just
calculated, there are two other key features of the density
fluctuations that follow as a consequence of slow-roll
single-field inflation.  The first is that the fluctuations are
adiabatic, which means that every component of the matter in the
universe --- the photons, the baryons, and the dark matter ---
fluctuate together.  The temperature can be related to the
density of photons, so it also fluctuates with the density
baryons or dark matter.  The reason for this feature is clear,
because the time delay affects all properties of the matter in
the universe the same way.  Until the perturbations reenter the
Hubble length (after which complicated things can happen), every
region of space behaves just like any other region of space,
except for a time offset.  Thus the matter content of any one
region can differ from that of some other region by at most an
adiabatic compression or expansion.  The WMAP
team~\cite{Komatsu10} has tested this relation for the
possibility of non-adiabatic fluctuations between photons and
cold dark matter.  By combining WMAP data with other data, they
find at the 95\% confidence level that the non-adiabatic
component is at most 6\% of the total in the case of
``axion-type'' perturbations, or 0.4\% in the case of
``curvaton-type'' perturbations.

The other key predication of slow-roll single-field inflation is
that the perturbations should be Gaussian.  Why are they
Gaussian?  They are Gaussian because $\delta \tilde \varphi (\vec
k,t)$ is calculated in a quantum field theory.  The perturbations
are small so we expect accurate results at lowest order, which
means that we are only calculating free-field-theory expectation
values, and they are Gaussian.  There are of course higher order
corrections, which in a given model can also be calculated, but
they are generically very small.  So, to first approximation, we
expect the answers to be Gaussian, which means in particular that
the three-point correlation function should vanish.  There has
been a lot of effort to look for non-Gaussianity, but so far no
convincing evidence for non-Gaussianity has been found.

\begin{figure}[t]
\begin{centering}
\epsfig{file=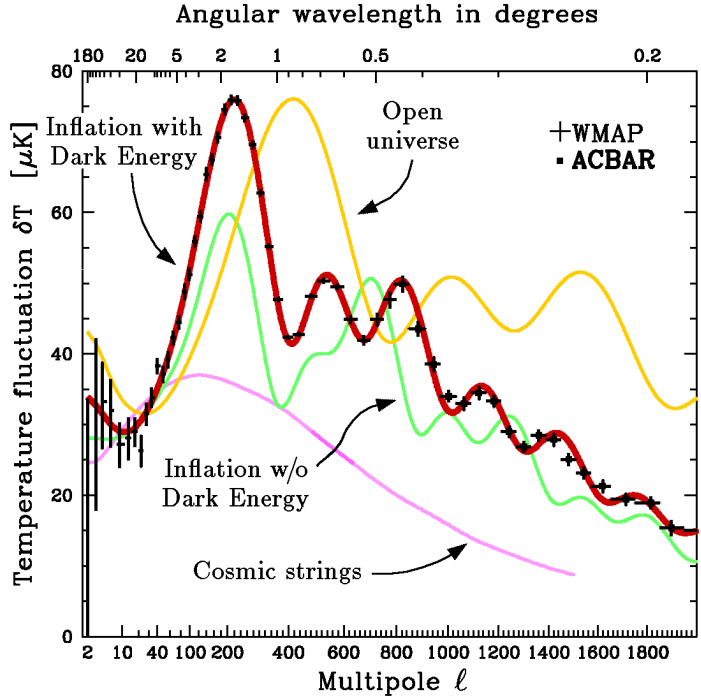}
\caption{Comparison of the latest observational measurements of
the temperature fluctuations in the CMB with several theoretical
models, as described in the text. The temperature pattern on the
sky is expanded in multipoles (i.e., spherical harmonics), and
the intensity is plotted as a function of the multipole number
$\ell$. Roughly speaking, each multipole $\ell$ corresponds to
ripples with an angular wavelength of $360^{\circ}/\ell$.}
\label{ahg:fig3}
\end{centering}
\end{figure}

The calculations shown here stop just after the end of
inflation, but with a lot of work by many astrophysicists the
calculations have been extended to make detailed predictions for
the fluctuations that can be detected today in the cosmic
microwave background.  The success is beautiful.  To process the
data, the temperature pattern observed in the CMB is expanded in
spherical harmonics, which is the spherical equivalent of Fourier
transforming, providing information about how the intensity of
the fluctuations varies with angular wavelength. 
Figure~\ref{ahg:fig3}~\cite{Tegmark12} shows the observed
temperature fluctuations as a function of the multipole number
$\ell$, using the 7-year WMAP data~\cite{Komatsu10} for $\ell <
800$, and ACBAR data~\cite{Reichardt09} for higher $\ell$. The red
line is the theoretical curve that comes about by extending the
inflationary predictions to the present day in a model with dark
energy ($\Lambda$) and cold dark matter, using the best-fit
parameters found by the WMAP team~\cite{Komatsu10}: ${\cal
P}_{\cal R}(0.002\ \hbox{Mpc}^{-1}) = 2.42 \times 10^{-9}$, $n_s
= 0.966$, $\Omega_\Lambda = 0.729$, $\Omega_{\rm dark\ matter}=
0.226$, $\Omega_{\rm baryon} = 0.045$, and $\tau = 0.085$, where
$\tau$ is the optical depth experienced by the photons since the
``recombination'' of the primordial plasma at about 380,000 years
after the big bang.  While there are 6 free parameters, 4 of them
have values that are expected on the basis of theory ($n_s
\approx 1$) or other observations ($\Omega_\Lambda$, $\Omega_{\rm
dark\ matter}$, and $\Omega_{\rm baryon}$), and they agree well.
One of the free parameters determines the overall height, so one
should not be impressed that the height of the primary peak
matches so well.  But the location, shape, and relative heights
of the peaks are really being predicted by the theory, so I
consider it a spectacular success.

For comparison, the graph also shows predictions for several
alternative theories, all of which are now ruled out by this
data.  The yellow line shows the expected curve for an open
universe, with $\Omega_{\rm total} = 0.30$.  The green line shows
an inflationary model with $\Omega_{\rm total} = 1$, but with
$\Omega_{\rm dark\ matter}=0.95$ and no dark energy.  The magenta
line shows the expectations for fluctuations generated by the
formation of cosmic strings in the early universe, taken from
Ref.~\refcite{Pen97}.  Structure formation caused by cosmic
strings or other ``defects'' was considered a viable possibility
before this data existed, but now cosmic strings are completely
ruled out as a major source of density fluctuations.

(There are possibly alternative ways to generate density
perturbations with the same properties as those of inflation, but
there is not yet a consensus about how easy it is to construct a
plausible model.  The cyclic ekpyrotic model~\cite{Khoury01,
Khoury01b, Tolley03, Khoury03, Khoury03b} was claimed to
naturally produce such fluctuations, but these claims were
disputed by a number of authors~\cite{Brandenberger01, Lyth01,
Tsujikawa02, Creminelli04}. Now at least some of the founders of
ekpyrosis~\cite{Khoury10,Khoury11} agree that the original models
do not give a nearly-scale invariant spectrum, as had been
claimed.  But these papers and others have proposed newer, more
sophisticated versions of bouncing universes, generally involving
either multiple fields, or settling for scale invariance for only
a limited range of scales.  Baumann, Senatore, and
Zaldarriaga~\cite{Baumann11} have argued that any single-field
model with attractor behavior has to be very close to de Sitter
space to remain weakly coupled for at least the required $\sim$10
$e$-folds needed to account for observations.)

\section{Outstanding Questions About Density
Perturbations}\label{ahg:sec6}

There are still a number of important, outstanding questions
concerning density perturbations:

\begin{arabiclist}
\item
Will $B$-modes be found?  Experiments are starting to measure the
polarization of the CMB, for which the spherical harmonic
expansion for the temperature pattern is replaced by an expansion
in $E$-modes and $B$-modes~\cite{Zaldarriaga96}. The $E$-modes are
those that can be expressed as gradients of scalar harmonic
functions, and they are produced as a by-product of the density
perturbations that we have been discussing.  The $B$-modes are
orthogonal to the $E$-modes; they cannot be expressed as
gradients of scalar modes, and they cannot be produced by density
perturbations.  There can be foreground contamination, but the
only known primordial source of $B$-modes is a background of
gravitational waves.  Thus, gravity waves might be discovered in
the CMB before they can be seen directly.  The discovery of a
primordial gravity wave background would be very exciting,
because it is the only thing that will give us a clue about the
energy scale at which inflation happened.  As far as we know now
inflation might have happened anywhere from the electroweak scale
up to the grand unified theory (GUT) scale, or a little beyond. 
The discovery of gravity waves would end the uncertainty, and
would also give strong evidence for the inflationary picture. 
There are, however, many inflationary models for which the energy
scale would be too low for the gravitational waves to be visible.

\item
Can sub-Planckian physics influence the calculation of
inflationary density perturbations?  A typical GUT-scale
inflationary model would include about 60 $e$-folds of inflation,
expanding by a factor of $e^{60} \approx 10^{26}$.  From the end
of inflation to today the universe would expand by another factor
of $\sim 10^{15}\ \hbox{GeV}/3\,\hbox{K} \approx 10^{27}$.  This
means that a distance scale of 1 m today corresponds to a length
of only about $10^{-53}$ m at the start of inflation, 18 orders
of magnitude smaller than the Planck length ($\sim 10^{-35}$ m). 
With a little more than the minimal amount of inflation --- which
would be a certainty in the eternal inflation picture to be
discussed below --- even the largest scales of the visible
universe would have been sub-Planckian at the start of inflation. 
So, it is relevant to ask whether inflation can possibly offer us
a glimpse of sub-Planckian physics.  There is of course no solid
answer to this question, since there is no real understanding of
how this process should be described.  Kaloper, Kleban, Lawrence,
and Shenker~\cite{Kaloper02} have argued that the perturbations
are determined primarily by local effective field theory on the
scale of order $H$, so that sub-Planckian effects would be
invisible except possibly in unconventional models for which the
fundamental string scale is many orders of magnitude below the
four-dimensional Planck mass, $\sim 10^{19}$ GeV. Some
authors~\cite{Groeneboom07, Avgoustidis12} have reached similar
conclusions, but other authors~\cite{Martin01, Kempf01, Easther01,
Collins09} have concluded that the effects might be much easier
to see.  The conclusions of Ref.~\refcite{Kaloper02} seem
plausible to me, but certainly the role of sub-Planckian physics
is not yet fully understood.

\item
Will effects beyond the single-field slow-roll approximation be
found?  With multiple fields, or with unusual features in the
potential for a single field, models can be constructed that
predict significant non-Gaussianity, non-adiabaticity, or
spectral distortions.  There is an active industry engaged in
studying models of this sort, and in looking for these
nonstandard features in the data.  The WMAP seven-year
analysis~\cite{Komatsu10} reports ``no convincing deviations from
the minimal model,'' but we all await the data from the Planck
mission, expected in less than a year, and the data from a
variety of ground-based experiments.

\end{arabiclist}

\section{Fluctuations on Larger Scales: Eternal
Inflation?}\label{ahg:sec7}

Since the density perturbation calculations have been incredibly
successful, it seems to make sense to take seriously the
assumptions behind these calculations, and follow them where they
lead.  I have to admit that there is no clear consensus among
cosmologists, but to many of us the assumptions seem to be
pointing to eternal inflation, and the multiverse. 

The mechanism for eternal inflation is described most efficiently
by separating the cases of the two types of potential functions
shown in Fig.~\ref{ahg:fig1}.  For the new inflation case, that
state for which the scalar field is poised on the top of the
potential hill is a metastable state, often called a ``false
vacuum,'' which decays by the scalar field rolling down the hill. 
This state decays exponentially, but in any working model of
inflation the half-life of the decay is much longer than the
doubling time associated with the exponential expansion.  Thus,
if we follow a region for a period of one half-life, at the end
of the period only half of the original region would be still be
inflating.  However, the half that is still inflating will have a
volume vastly larger than the volume of the entire region at the
start, so the process will go on forever.  Each decay will lead
to the production of a ``pocket'' universe, and the creation of
pocket universes will go on forever, as pieces of the
ever-growing false vacuum region undergo decays.  Once inflation
starts, it never stops.%
\footnote{The first models of eternal new inflation were proposed
by Steinhardt~\cite{Steinhardt83} and Linde~\cite{Linde82b}. 
Vilenkin~\cite{Vilenkin83} was the first to describe eternal
inflation as a generic feature of new inflation.}

For the case of a chaotic-type potential, as in
Fig.~\ref{ahg:fig1}(b), naively one would think that the field
would inexorably roll down the hill in some finite amount of
time.  However, Linde~\cite{Linde86} discovered that when quantum
fluctuations are taken into account, this need not be the case. 
To understand this, consider an inflating region of space of size
$H^{-1}$, with the inflaton field $\varphi$ approximately uniform
over this region, at some value $\varphi_0$.  After one Hubble
time ($H^{-1}$) the region will have expanded by $e^3 \approx
20$, and can be viewed as 20 Hubble-sized regions which will
start to evolve independently.  The average field $\varphi$ in
any one of these regions will usually be lower than $\varphi_0$,
due to the classical rolling down the hill, but the classical
evolution will be modified by random quantum jumps, which can be
estimated as $\sim H/(2 \pi)$.  It is therefore possible that in
one or more of these 20 regions, $\varphi$ can equal or exceed
$\varphi_0$.  A back-of-the-envelope calculation shows that if
\begin{equation}
  {H^2 \over |\dot \varphi |} \simgt 5 \ ,
\label{ahg:eq41}
\end{equation}
then the expectation value for the number of regions with
$\varphi > \varphi_0$ is greater than one.  That implies that the
number of Hubble-sized regions with $\varphi > \varphi_0$ will
grow exponentially with time, and the inflation becomes eternal. 
Note that $H^2 / |\dot \varphi| \approx \sqrt{{\cal P_{\cal R}}}
\approx (G V)^{3/2}/|V'|$, so the eternally inflating behavior is
really the large-$\varphi$, long-wavelength, tail of the density
perturbation spectrum.  Since $V^{3/2}/|V'|$ grows without bound
as $\varphi \to \infty$ for most potentials under consideration,
almost all models allow for eternal inflation.

There is certainly no proof that we live in a multiverse, but I
will argue that there are three winds --- that is, three
independent scientific developments, arising from three different
branches of science --- which seem to be leading to the
multiverse picture. 

\begin{arabiclist}
\item
{\it Theoretical Cosmology: Eternal Inflation.}  As I just
described, almost all inflationary models are eternal into the
future.

\item
{\it String Theory: The Landscape.} String theory predicts that
there is not just one kind of vacuum, but instead there are a
colossal number of them: $10^{500}$ or maybe more~\cite{Bousso00,
Susskind03}. The underlying laws of physics would be the same
everywhere, but nonetheless each type of vacuum would create an
environment in which the low-energy laws of physics would be
different.  Thus, if there is a multiverse, it would be a varied
multiverse, in which the different pocket universes would each
appear to have their own laws of physics.

\item
{\it Observational Astronomy: the Cosmological Constant.} The
third ``wind'' has its roots in the fine-tuning that our universe
appears to exhibit.  In the past a minority of physicists argued
that things such as the properties of ice or the energy levels of
carbon-12 appeared to be fine-tuned for the existence of life,
but not very many scientists found this convincing.  If these
properties were different, then maybe life would form some other
way.  However, a form of fine-tuning that many of us find much
more convincing became evident starting in 1998, when two groups
of astronomers~\cite{Riess98, Perlmutter98} announced that the
expansion of the universe is not slowing down due to gravity, but is
in fact accelerating.  The simplest explanation is that the
acceleration is caused by a nonzero energy density of the vacuum,
also known as a cosmological constant.  But that would mean that
the vacuum energy density is nonzero, yet a full 120 orders of
magnitude smaller than the Planck scale ($M_{\rm Pl}^4$, where
$M_{\rm Pl} = 1/\sqrt{G}$), the scale that most theoretical
physicists would consider natural.  Physicists have struggled to
find a physical explanation for this small vacuum energy density,
but no generally accepted solution has been found.  But if the
multiverse is real, the problem could go away.  With $10^{500}$
different types of vacuum, a small fraction, but nonetheless a
large number of them, would be expected to have an energy density
as small as what we observe.  The smallness of the vacuum energy
density would be explained, therefore, if we could explain why we
should find ourselves in such an unusual part of the multiverse. 
But as pointed out by Weinberg and
collaborators~\cite{Weinberg87, Martel97} some time ago, there is
a selection effect.  If we assume that life requires the
formation of galaxies, then one can argue that life in the
multiverse would be concentrated in those pocket universes with
vacuum energy densities in a narrow band about zero.  Thus, while
a typical vacuum energy density in the multiverse would be on the
order of the Planck scale, almost all life in the multiverse
would find a small value, comparable to what we see.

\end{arabiclist}

\medskip

While the multiverse picture looks very plausible in the context
of inflationary cosmology --- at least to me --- it raises a
thorny and unsolved problem, known as the ``measure problem.''
Specifically, we do not know how to define probabilities in the
multiverse.  If the multiverse picture is right, then anything
that can happen will happen an infinite number of times, so any
distinction between common and rare events requires the
comparison of infinities.  Such comparisons are not
mathematically well-defined, so we must adopt a recipe, or
``measure,'' to define them.  Since the advent of quantum theory
essentially all physical predictions have been probabilistic,
so probability is not a concept that we can dispense with.  To
date we do not understand the underlying physical basis for such
a measure, but much progress has been made in examining proposals
and ruling out many of them.%
\footnote{For a recent summary of a field that is in
a state of flux, see Ref.~\refcite{Freivogel11}.  For a new
proposal that was advanced since this summary, see
Ref.~\refcite{Nomura11}.}

One might guess that this problem is easily handled by choosing a
finite sample spacetime region in the multiverse, calculating the
relative frequencies of different types of occurrences in the
sample region, and then taking the limit as the region becomes
infinite.  This seems like a very reasonable approach, and in
fact most of the measure proposals that have been discussed are
formulated in this way.  The problem is that the answers one
obtains are found to depend sensitively on the method that is
used to choose the sample region and to allow it to grow.  The
dependence on the method of sampling seems surprising, but it
sounds plausible if we remember that the volume of the multiverse
grows exponentially with time.  A sample spacetime region will
generally have some final time cutoff, and the spacetime volume
will generally grow exponentially with the cutoff.  But then, no
matter how large the cutoff is taken, the volume will always be
dominated by a region that is within a few time constants of the
final time cutoff hypersurface.  No matter how large the final
time cutoff is taken, the statistics will never be dominated by
the interior of the sample region, but instead instead will be
dominated by the final time cutoff surface.  For that reason, it
is not surprising that the method of choosing this surface will
always affect the answers.

There are a number of important questions, in the multiverse
picture, that depend very crucially on the choice of measure. 
How likely is it that we observe a vacuum energy density as low
as what we see?  How likely is it that our universe has a mass
density parameter $\Omega$ sufficiently different from 1 that we
can hope to measure the difference?  How likely is it that we
might find evidence that our pocket universe collided with
another sometime in its history?  And, if there are many vacua in
the landscape of string theory with low energy physics consistent
with what we have measured so far, how likely is it that we will
find ourselves in any particular one of them?  If we live in a
multiverse, then in principle all probabilities would have to be
understood in the context of the multiverse, but it seems
reasonable to expect that any acceptable measure would have to
agree to good accuracy with calculations that we already know to
be successful.

As discussed in Ref.~\refcite{Freivogel11}, we can identify a
class of measures that give reasonable answers.  It seems
plausible that the ultimate solution to this problem will give
similar answers, but the underlying principles that might
determine the right answer to this question remain very
mysterious.  Nonetheless, the success of inflation in explaining
the observed properties of the universe, including the density
perturbation predictions discussed here, provides strong
motivation to expect that some solution to the measure problem
will be found.

\begin{acknowledgments}
This work is supported in part by the U.S. Department of Energy (DOE)
under cooperative research agreement DE-FG02-05ER41360.
\end{acknowledgments}

\end{document}